\documentclass{pasj01}

\RequirePackage{mathpazo}
\RequirePackage[T1]{fontenc}

\newcounter{author}
\def\authorcount#1#2{\refstepcounter{author}\label{#1}
                     \altaffiltext{\ref{#1}}{#2}}

\begin{document}

\SetRunningHead{A.Imada et al.}{}

\Received{201X/XX/XX}
\Accepted{201X/XX/XX}

\title{On the Colour Variations of Negative Superhumps}

\author{Akira~\textsc{Imada},\altaffilmark{\ref{affil:hamburg}}$^,$\altaffilmark{\ref{affil:kwasan}*}
Kenshi~\textsc{Yanagisawa},\altaffilmark{\ref{affil:oao}}
Nobuyuki~\textsc{Kawai}\altaffilmark{\ref{affil:ttec}}
}


\authorcount{affil:hamburg}{Hamburger Sternwarte, Universit\"at Hamburg, Gojenbergsweg 112, D-21029 Hamburg, Germany}

\authorcount{affil:kwasan}{
  Kwasan and Hida Observatories, Kyoto University, Yamashina, Kyoto 607-8471, Japan}
\email{$^*$a\_imada@kusastro.kyoto-u.ac.jp}


\authorcount{affil:oao}{
  Okayama Astrophysical Observatory, National Astronomical Observatory of Japan, Asakuchi, Okayama 719-0232, Japan}

\authorcount{affil:ttec}{
Department of Physics, Tokyo Institute of Technology, Ookayama 2-12-1, Meguro-ku, Tokyo 152-8551, Japan}


 
\maketitle

\begin{abstract}
We present simultaneous $g'$, $R_{\rm c}$, and $I_{\rm c}$ photometry of the notable dwarf nova ER UMa during the 2011 season. Our photometry revealed that the brightness maxima of negative superhumps coincide with the bluest peaks in $g' - I_{\rm c}$ colour variations. We also found that the amplitudes of negative superhumps are the largest in the $g'$ band. These observed properties are significantly different from those observed in early and positive superhumps. Our findings are consistent with a tilted disk model as the light source of negative superhumps.

\end{abstract}

\section{Introduction}

ER UMa-type dwarf novae are a subclass of SU UMa-type dwarf novae and their main characteristics are that their supercycle is unusually short, typically less than 50 d \citep{kat95eruma}. ER UMa-type dwarf novae are also known as the systems exhibiting negative superhumps in their light curves (\cite{pat95v1159ori}; \cite{ohs14eruma}), and are considered to be a result of retrograde precession of a tilted disk \citep{woo07negSH}. During appearance of negative superhumps, the importance of gas stream overflow was discussed by \citet{bar88tvcol}, in which they suggested that the physical source of the negative superhump period was the sweeping of the accretion stream hot spot across the face of a tilted disk. \citet{mon09negativeSH} performed SPH simulations and suggested that the light source of negative superhumps is originated from the innermost disk annuli, where gas stream overflow plays a crucial role. Overall, the light curves of negative superhumps are believed to be caused by changes in the location of the hot spot where the gravitational energy is released.

In order to understand the unusual properties of ER UMa-type dwarf novae, extensive photometric and spectroscopic studies, not only during outburst, but also during quiescence, have been performed since the discovery of this subclass (\cite{rob95eruma}; \cite{gao99erumaSH}). In particular, \citet{ohs12eruma} discovered negative superhumps during the 2011 January superoutburst of ER UMa itself. At present, this is the sole case in which negative superhumps have been observed during superoutburst of ER UMa-type dwarf novae. This discovery has motivated us to closely monitor the object.

Although many authors performed extensive CCD photometry via a worldwide campaign (e.g., \cite{VSNET}), most of the data was acquired without the use of filters. On the other hand, multicolour photometry has revealed the drastic colour variations of early and positive superhumps \citep{mat09v455and}. More recently, \citet{ima18hvvirj0120} reported that the light source of early superhumps is cool and vertically-expanded at the outer region of the accretion disk, based on simultaneous optical and near-infrared photometry. As for the positive superhumps, \citet{neu17j1222} showed that the brightness maximum differs from the bluest peak of $B - I$ colour variations. This result suggests that the pressure effect in the accretion disk plays a role. Although multicolour photometry of early and positive superhumps has been reported over the past decade, we have less information on colour variations of negative superhumps.

In this letter, we report on the colour variations of negative superhumps in ER UMa for the first time. Our multicolour photometry has revealed that the brightness maxima of negative superhumps coincide with the bluest peaks in the $g' - I_{\rm c}$ colour variations. Also, the $g'$ band shows the largest amplitude when negative superhumps appear. These phenomena are in contrast with those observed in early and positive superhumps. Detailed photometric studies will be published in a forthcoming paper.

\section{observations and analyses}

Time-resolved CCD photometry was performed from 2011 February 3 to 2011 October 9 using the MITSuME 50cm-telescope located in Okayama Astrophysical Observatory.\footnote{www.oao.nao.ac.jp} The MITSuME telescope can obtain $g'$, $R_{\rm c}$, and $I_{\rm c}$ bands simultaneously, which allows us to study magnitudes and colour variations of variable stars \citep{3me}. Detailed observational log will be published in a forthcoming paper. All of the observational data were acquired with a 30-sec exposure. The images obtained were processed with the standard IRAF/daophot software.\footnote{IRAF (Image Reduction and Analysis Facility) is distributed by the National Optical Astronomy Observatories, which is operated by the Association of Universities for Research in Astronomy, Inc., under cooperative agreement with the National Science Foundation.}
We adopted differential photometry using the star at RA: 09$^{h}$47$^{m}$14$^{s}$.64, DEC: +51$^{\circ}$48'40''.3 (epoch: J2000), $g'$ = 12.698, $R_{\rm c}$ = 11.898, and $I_{\rm c}$ = 11.402 
 as a comparison star, whose constancy was checked against nearby stars in the same image. The total datapoints during our campaign exceeded 35000, which is the most extensive multicolour observations for ER UMa to date. All of the data were converted to the mid-exposure times of the Barycentric Julian Date (BJD).

\section{Results}

\begin{figure*}
\begin{center}
\FigureFile(160mm,80mm){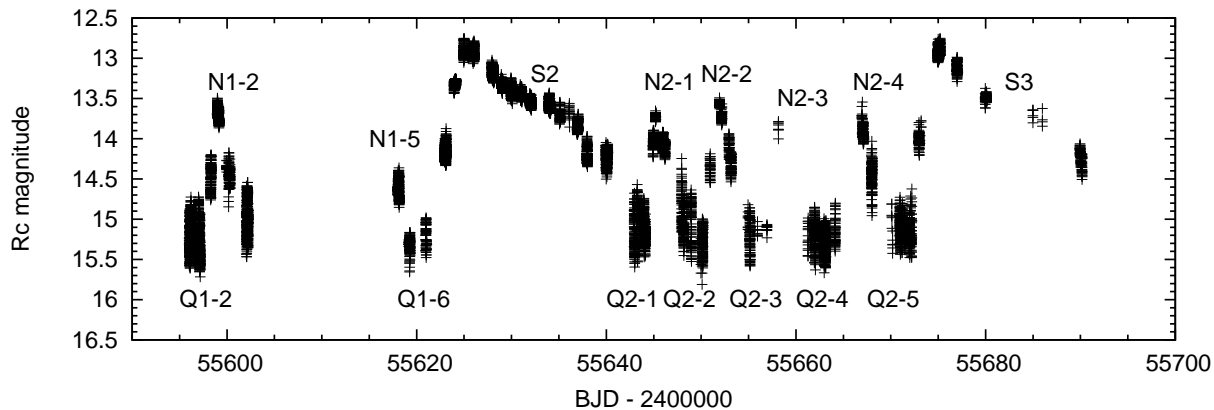}
\end{center}
\caption{A part of $R_{\rm c}$ light curve of ER UMa in our campaign. The names of the outbursts are taken from those designated in \citet{ohs14eruma}. We named each quiescence (abbreviated as Q), normal outburst (abbreviated as N), and superoutburst (abbreviated as S) as described here. As noted in \citet{ohs14eruma}, negative superhumps were visible in the most of quiescent light curves.}
\label{lc}
\end{figure*}

\begin{figure*}
\begin{center}
\FigureFile(160mm,160mm){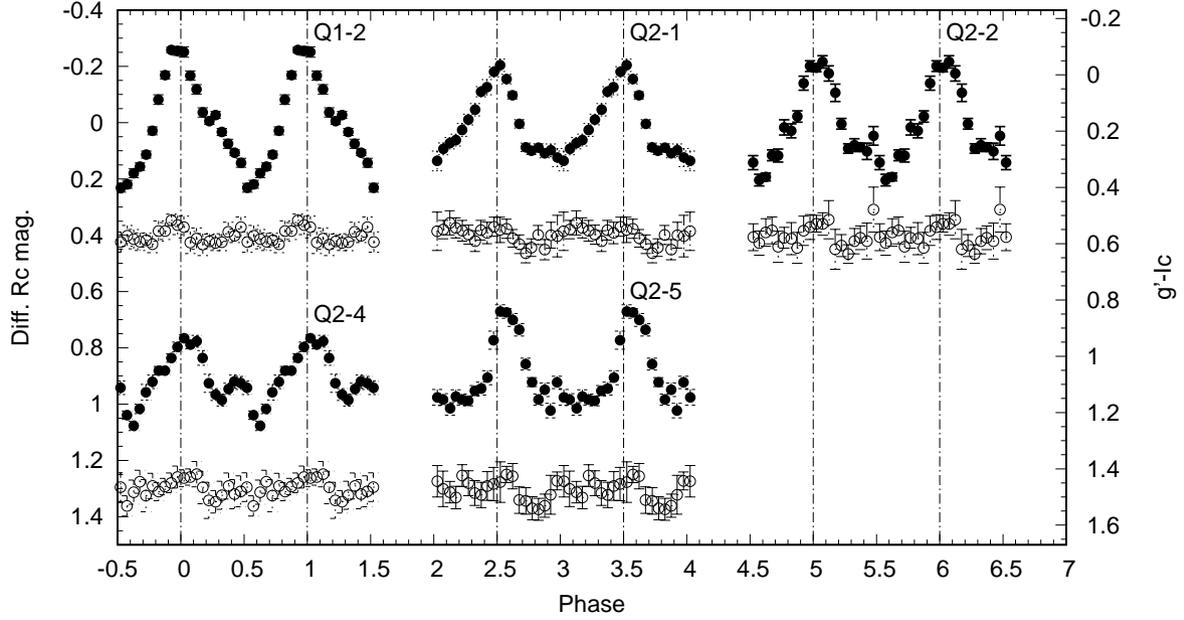}
\end{center}
\caption{Phase-averaged $R_{\rm c}$ light curves (filled circles) and $g' - I_{\rm c}$ colour variations (open circles) of negative superhumps folded with the periods estimated with the PDM method. The vertical-left axis denotes the differential $R_{\rm c}$ magnitude, and the vertical-right axis denotes $g' - I_{\rm c}$ colour variations. The differential $R_{\rm c}$ magnitude applies to the phase-averaged $R_{\rm c}$ light curve in figure \ref{lc}. For better visualization, phase-averaged $R_{\rm c}$ light curve and $g' - I_{\rm c}$ colour variations are repeated twice in phase for each Q segment. As for the phase-averaged light curves of Q2-4 and Q2-5 segments, they are shifted by 1 in the differential $R_{\rm c}$ magnitude. We also shifted $g' - I_{\rm c}$ colour variations of Q2-4 and Q2-5 by ${\sim}$ 0.8. Note that the bluest peaks of $g' - I_{\rm c}$ colour variations coincide with brightness maxima of $R_{\rm c}$ band.}
\label{neg}
\end{figure*}

\begin{figure}
\begin{center}
\FigureFile(80mm,60mm){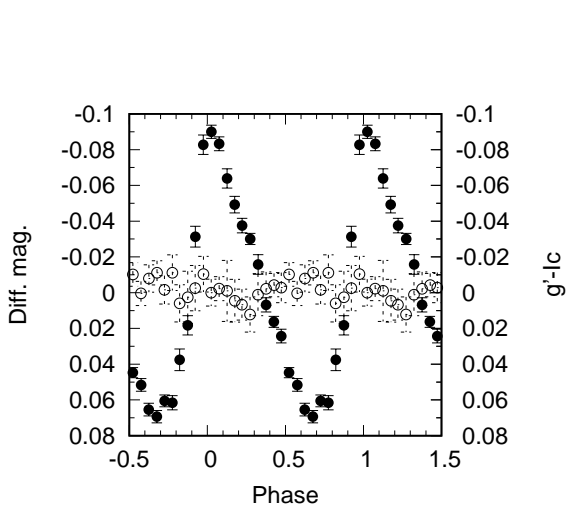}
\end{center}
\caption{Phase-averaged $R_{\rm c}$ light curves (filled circles) and $g' - I_{\rm c}$ colour variations (open circles) on BJD 2455628, corresponding to the early phase of the stage B during S2 in figure 1. The vertical-left axis denotes the differential $R_{\rm c}$ magnitude, and the vertical-right axis denotes $g' - I_{\rm c}$ colour variations. The differential $R_{\rm c}$ magnitude applies to the phase-averaged $R_{\rm c}$ light curve in figure \ref{lc}. The data were folded with 0.06592 d, corresponding to the strongest signal of the positive superhump period during the stage B of S2. Note that the bluest peak in $g' - I_{\rm c}$ colour variations occurs around phase 0.6-0.8 and differ from the maximum magnitude of positive superhumps.}
\label{pos}
\end{figure}

Figure \ref{lc} shows the representative $R_{\rm c}$ light curve of our observations, in which two superoutbursts and six normal outbursts are visible. The names of each outburst use the same description as those in \citet{ohs14eruma}. Here we define the names of quiescence as those described in their figure. Although we are not confident to detect negative superhumps in Q1-6 and Q2-3 because of insufficient data, the other quiescent light curves showed conspicuous signals of negative superhumps with periods of 0.06249(5) d (Q1-2), 0.06234(9) d (Q2-1), 0.06249(4) d (Q2-2), 0.06256(9) d (Q2-4) and 0.06236(8) d (Q2-5), by applying the Phase Dispersion Minimization method (PDM, \citet{pdm}). We use these periods in the following analyses. These periods are shorter than the orbital period of ER UMa ($P_{\rm orb}$ = 0.06366 d, \cite{tho97erumav1159ori}). The errors of the PDM method were estimated using the Lafter-Kinmann class method developed by \citet{fer89error} and \citet{pdot2}. Also, a hint of negative superhumps was observed near the end of S2, as pointed out by \citet{ohs14eruma}.
Figure \ref{neg} shows the phase-averaged $R_{\rm c}$ light curve and $g' - I_{\rm c}$ colour variations, folded with the periods obtained above. As can be seen in these figures, the brightness maxima of negative superhumps are associated with the bluest peaks in the $g' - I_{\rm c}$ colour variations. We obtained the correlation coefficient between $R_{\rm c}$ and $g' - I_{\rm c}$ colours variations to be 0.49 (Q1-2), 0.34 (Q2-1), 0.41 (Q2-2), 0.71 (Q2-4), and 0.35 (Q2-5). In figure \ref{pos}, we also present phase-averaged $R_{\rm c}$ and $g' - I_{\rm c}$ colour variations of positive superhumps observed on BJD 2455627.92$-$28.12 folded with 0.06592(5) d, corresponding to the strongest signal in $R_{\rm c}$ band during the stage B of S2 using the PDM method.\footnote{Period changes of positive superhumps during superoutburst consist of three stages (stage A, B, and C). For a detailed characteristics of each stage, see \citet{Pdot}.} \citet{ohs14eruma} reported that positive and negative superhumps coexisted during S2 except for the early stage of the superoutburst. However, we were unable to detect a hint of negative superhumps on BJD 2455627.92$-$28.12, possibly due to the short coverage of our data and a much weaker signal of negative superhumps than positive ones. Although $g' - I_{\rm c}$ colour variations are weak, one can see a tendency that $g' - I_{\rm c}$ are blueish at phase 0.6-0.8, corresponding to the brightness minima of positive superhumps. During BJD 2455627.92$-$28.12 when the positive superhumps appeared, we obtained the correlation coefficient between $R_{\rm c}$ and $g' - I_{\rm c}$ colour variations to be -0.34. It should be noted that the correlation coefficient during positive superhumps is significantly different from that during negative superhumps.

\begin{figure}
\begin{center}
\FigureFile(80mm,60mm){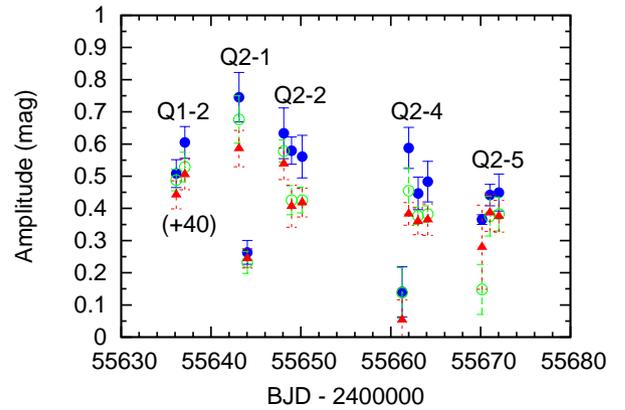}
\end{center}
\caption{Nightly-averaged amplitudes of negative superhumps during quiescence. Filled circles (blue), open circles (green), and filled triangles (red) denote the amplitudes of $g'$, $R_{\rm c}$, and $I_{\rm c}$ bands, respectively. The datapoints of Q1-2 are shifted by 40 d for better visualization. Note that the $g'$ band show the largest amplitudes.}
\label{amp}
\end{figure}

We also examined nightly-averaged amplitudes of negative superhumps, which display in figure \ref{amp}. In order to derive the amplitudes in each band, we folded nightly-averaged $g'$, $R_{\rm c}$, and $I_{\rm c}$ light curves with the above-obtained negative superhump period in each quiescent light curve, and measured these amplitudes. As reported in \citet{ohs14eruma}, the amplitudes of negative superhumps were as large as ${\sim}$ 0.5 mag except for a few nights. It should be noted that the amplitudes of $g'$ band are the largest. This property is opposite to that observed in early superhumps, in which redder passbands show larger amplitudes \citep{ima18hvvirj0120}. We derived the mean amplitudes of negative superhumps in $g'$, $R_{\rm c}$, and $I_{\rm c}$ to be 0.49, 0.40, and 0.38 (in units of magnitude), respectively.

\section{Discussion}

The present observations of ER UMa have revealed new characteristics of negative superhumps: the brightness maxima of negative superhumps coincide with the bluest peaks in $g' - I_{\rm c}$ colour variations, and the $g'$ band has the largest amplitudes.

In the previous works concerning colour variations of positive superhumps, many authors reported that there is a difference in phase between the bluest peak of $g' - I_{\rm c}$ colour variations and brightness maximum of positive superhumps. For example, \citet{mat09v455and} reported that the bluest peaks in $g' - I_{\rm c}$ colour variations are prior to the brightness maxima by a phase of 0.15 during the 2007 superoutburst of V455 And. On the other hand, \citet{iso15ezlyn} performed $g'$ and $i'$ band photometry of the 2010 superoutburst of EZ Lyn and found that the positive superhump amplitude is anticorrelated with the $g' - i'$ colour, which means that the $g' - i'$ peak is prior to the positive superhump maximum by a phase of ${\sim}$ 0.5. Similar observations were reported in other WZ Sge-type dwarf novae (\cite{nak13j0120}; \cite{neu17j1222}; \cite{ima18hvvirj0120}).  \citet{mat09v455and} noted that the phase difference between the bluest peak of colour variations and the positive superhump maximum can be interpreted as the time scale of the expansion of the positive superhump light source. Recently, \citet{neu17j1222} studied the relation between colour variations and magnitudes of positive superhumps according to the superoutburst stages. They found that growing superhumps (during stage A) showed weak colour variations while superhumps with increasing periods (during stage B) showed large colour variations. \citet{ima18hvvirj0120} reached the same conclusion as \citet{neu17j1222}. \citet{ima18hvvirj0120} suggested that the large $g' - I_{\rm c}$ colour variations may reflect the strong pressure effect in stage B. It is likely that the phase difference between colour and magnitude can be caused by the change in temperature and pressure in the accretion disk.

Taking the above statements into consideration, the present finding may imply that the accretion disk has lower pressure and temperature effects during the appearance of negative superhumps. From the theoretical side, \citet{mon10negSH} modeled the accretion disk in which the spiral structures in the inner region play a role in generating negative superhumps. With regard to the spiral structures, the evidence for them was also provided during the appearance of early superhumps in WZ Sge \citep{bab02wzsgeletter}. As noted above, early superhumps show the difference in phase between maximum magnitude and colour variations. In such a condition, the drastic variations of the global disk structure is expected. Our finding suggests that the variations of the inner disk structures, such as variation of the spiral structures, play a minor role in the observed light curves of negative superhumps.

\citet{ima18hvvirj0120} studied amplitudes of early and positive superhumps. Early superhumps show the largest amplitude in the $K_{\rm s}$ band and the smallest amplitude in the $g'$ band. Based on the observations, \citet{ima18hvvirj0120} argued that the light source of early superhumps is generated in the outer region of the vertically-expanded accretion disk. As for the positive superhumps, \citet{ima18hvvirj0120} reported that these provide no evidence for dependence on wavelength. In combination with the statistical study of the amplitudes of positive superhumps according to the system inclination \citep{pdot3}, it may be that the light source of positive superhumps is geometrically thin \citep{ima18hvvirj0120}. In the case of the amplitudes of negative superhumps in ER UMa, the largest amplitude in the $g'$ band implies that the light source of negative superhumps may have originated from the inner region of the accretion disk.

In summary, the present findings suggest that the light source of negative superhumps is associated with the inner region of the accretion disk, but the drastic variations of temperature and pressure, such as observed in early and positive superhumps, are almost absent when negative superhumps appear. The present findings are likely to be consistent with a tilted disk model as the light source of negative superhumps \citep{woo07negSH}. The lack of pressure in the inner disk may further support the theoretical study of \citet{mon09negativeSH} in which she states that the gas stream overflow in the inner tilted disk is indispensable in generating negative superhumps. Because the gas stream overflow is observed in many types of binary systems such as X-ray binary systems and protostellar systems, we suggest that our findings can also apply to these systems. Also, theoretical studies show that the relation between the positive superhump excess (${\epsilon}_{\rm +}^{*}$) and negative superhump deficit (${\epsilon}_{\rm -}^{*}$) is described as ${\epsilon}_{\rm +}^{*}$/|${\epsilon}_{\rm -}^{*}$| ${\sim}$ 7/4 in the absence of the pressure effect (\cite{lar98XBprecession}; \cite{osa13v344lyrv1504cyg})\footnote{${\epsilon}^{*}$ ${\equiv}$ 1 - $P_{\rm orb}$/$P_{\rm sh}$}. Empirically, this value is known to be close to 2 (\cite{pat97v603aql}; \cite{woo09negativeSH}). However, some systems significantly deviate from this trend. For example, \citet{kat13j1924} reported that this value was close to 1 for KIC 8751494. \citet{kat13j1924} noted that this deviation can be understood if the pressure effect is taken into consideration. We expect that the phase difference between the negative superhump magnitude and the $g' - I_{\rm c}$ colour variations can be observed for systems in which ${\epsilon}_{\rm +}^{*}$/|${\epsilon}_{\rm -}^{*}$| deviates from 7/4. This should be investigated in future multicolour photometry in various systems exhibiting negative superhumps.



\begin{ack}
We would like to thank an anonymous referee for helpful comments on the manuscript of the letter. We acknowledge with thanks the variable star observations from the AAVSO and VSNET International Database contributed by observers worldwide and used in this research. We would like to express our gratitude to Roger D. Pickard and Daisaku Nogami for helpful comments on the manuscript of the letter. This work is partly supported by the Publication Committee of the National Astronomical Observatory of Japan (NAOJ).
\end{ack}

\bibliographystyle{pasjtest1}
\bibliography{cvs2016}

\end{document}